\documentclass[amsmath,amssymb,aip,reprint]{revtex4-1}

\usepackage{amsfonts}
\usepackage{amsthm}
\usepackage{graphicx}
\usepackage{hyperref}
\usepackage{bbm}
\usepackage[utf8]{inputenc}
\usepackage[T1]{fontenc}
\usepackage{mathptmx}
\usepackage{etoolbox}

	\newcommand{\ncd}{\newcommand}
	\ncd{\mrm}    {\mathrm}
	\ncd{\beq} {\begin{equation}}
	\ncd{\eeq} {\end{equation}}

	\def\d{{\rm d}}

	\newcommand{\re}{\mathbb{R}}
	\newcommand{\B}{\mathcal{B}}
	
	\newcommand{\lie}{\mathcal{L}}

	\newcommand{\prt}[1]{\frac{\partial}{\partial #1}}

	\newcommand{\bea}{\begin{align}}
	\newcommand{\eea}{\end{align}}
	\newcommand{\eq}[1]{(\ref{#1})}

\usepackage{tikz}
\usetikzlibrary{decorations.pathreplacing,arrows,positioning,angles,calc,external}
	\tikzset{
    		>=stealth',pil/.style={->,thick,shorten <=2pt,shorten >=2pt,}
		}
 \usepackage{pgfplots}
 	\pgfplotsset{compat=1.3,every axis/.append style={font=\small,thin,tick style={ultra thin}}}
  	\usepgfplotslibrary{fillbetween}

\makeatletter
\def\@email#1#2{%
 \endgroup
 \patchcmd{\titleblock@produce}
  {\frontmatter@RRAPformat}
  {\frontmatter@RRAPformat{\produce@RRAP{*#1\href{mailto:#2}{#2}}}\frontmatter@RRAPformat}
  {}{}
}%
\makeatother

\begin{document}

	\title{Light propagation through optical media using metric contact geometry}
	
		\author{D. Garc\'ia-Pel\'aez}
		\email{dgarciap@up.edu.mx}
		\affiliation{Universidad Aut\'onoma Metropolitana Azcapotzalco,\\ Avenida San Pablo Xalpa 180, Azcapotzalco, Reynosa Tamaulipas, 02200 Ciudad de 
		M\'exico, M\'exico}
		\affiliation{Universidad Panamericana, Tecoyotitla 366.
Col. Ex Hacienda Guadalupe Chimalistac, 01050 M\'exico D.F., M\'exico}

		\author{C. S. L\'opez-Monsalvo}
		\affiliation{Universidad Aut\'onoma Metropolitana Azcapotzalco,\\
		Avenida San Pablo Xalpa 180, Azcapotzalco, Reynosa Tamaulipas, 02200 Ciudad de 
		M\'exico, M\'exico}
		
		\author{A. Rubio Ponce}
		\affiliation{Universidad Aut\'onoma Metropolitana Azcapotzalco,\\ Avenida San Pablo Xalpa 180, Azcapotzalco, Reynosa Tamaulipas, 02200 Ciudad de 
		M\'exico, M\'exico}
		\affiliation{Departamento de F\'isica, Centro de Investigaci\'on y de Estudios Avanzados del Instituto Polit\'ecnico Nacional,\\ AP 14740, CP 07300, Ciudad de M\'exico, M\'exico}

\date{\today}

	\begin{abstract}

In this work, we show that the orthogonality between rays and fronts of light propagation in a medium is expressed in terms of a suitable metric contact structure of the optical medium without boundaries. Moreover, we show that considering interfaces (modeled as boundaries) orthogonality is no longer fulfilled, leading to optical aberrations and in some cases total internal reflection.  We present some illustrative examples of this latter point. 



	\end{abstract}

\maketitle

\section{Introduction}

Geometric optics has acquired a renewed interest in theoretical and applied physics.
 The inception of geometric techniques and methods to control the path of light inside a medium provided us with the foundations of a new material science. Arising in the context of gravitational lensing, the geometric studies regarding Fermat's and Huygens' principles brought a new perspective in the understanding and modelling of wave phenomena in non-trivial media.

Soon after the advent of general relativity, it was observed that the gravitational field \emph{bends} the path followed by a light ray. This was confirmed by Eddington after noticing that the apparent position of the stars are shifted from their expected position in the sky when observed during a solar eclipse. This \emph{bending} phenomenon is analogous to the deviation of light rays while traversing  a medium whose refractive index changes from one place to the other. In this sense, it was shown that a  class of optical media can be modeled by means of  a metric tensor  encoding its electromagnetic properties \cite{gordon1923lichtfortpflanzung,de1971gravitational,ehlers2012republication}. Furthermore, this analogy has evolved into the very active field of transformation optics, where the techniques and tools of differential geometry -- that have been insightful in gravitational physics -- have found its way into the more applied area of material science \cite{pendry2006controlling,leonhardt2006general,chen2010transformation,LOPEZMONSALVO2020168270}. In addition,  this has also been used in modelling analogue gravitational spacetimes such as  black holes  and cosmological solutions \cite{PhysRevA.102.023528,schuster2019electromagnetic,faccio2013analogue,schuster2018bespoke}, or to approximate curved spaces by $N$-dimensional \emph{simplices} in order to describe light propagation on them \cite{GeorgantzisGarcia:20}.

Fermat's principle poses  a well known problem in the calculus of variations.  In the non-relativistic setting -- where the notion of time is \emph{absolute} and \emph{universal} -- it states that light travels between two given \emph{spatial} locations following a path minimising a time functional. Such a perspective is clearly untenable in the context of General Relativity, where  it is commonly replaced by the assumption that the path taken by light in traveling from one location to another corresponds to a null geodesic. However, albeit it remains a variational problem, its precise formulation is far more elaborate (see Theorem 7.3.1 in \cite{perlick2000ray}).  

Similarly (cf. Theorem 7.1.2 in \cite{perlick2000ray}), Huygens' principle is centered in  the idea of light emissions being simultaneous for different observers at each moment in time \cite{berest1994huygens}. In this way,  light emission is described by well localised wavefronts, as they travel through  three dimensional space. In the relativistic setting,  the electromagnetic field at a precise event in spacetime  depends  on initial conditions originated its past null cone \cite{harte2013tails}. This has led to explore different wave phenomena where Huygens' principle is not fulfilled, most of which are due to dissipation processes where the wave distribution has not converging tails.

In recent years,  contact geometry has become  a unifying framework for various physical theories, eg. it provides a solid foundation for thermodynamics, non-conservative classical mechanics, electromagnetic fields among other applications \cite{kholodenko2013applications,garcia2014infinitesimal,bravetti2015contact,bravetti2017contact,lopez2021contact,flores2021contact, PhysRevLett.127.061102}. In particular, it has been used to exhibit an explicit correspondence between  Fermat's and Huygens' principles  \cite{HGeiges-ABHCGandT}. In this sense, the aim of this work is to explore the use of  contact transformations in the description of light propagation in \emph{optical media}. 

Here we assume that an optical medium can be represented by a Riemannian manifold $(\B,g)$ where $\B$ is considered to be the space occupied by a material embedded in spacetime while $g$ is its corresponding  optical  metric. Moreover, the geodesic flow in its unitary tangent bundle can be represented by a contact transformation acting on its space of contact elements. This fact allows us to describe the wavefront evolution in an optical medium solely in terms of the contact transformation and to  reconstruct the geodesic flow from the Reeb vector field. This provides us with  a way of constructing wavefronts in optical media without directly solving the wave equation. This approach is particularly useful to explore the relationship between rays and fronts as they move across interfaces. While this is simply expressed as Snell's Law in homogeneous materials, we will show that this construction helps in visualising phenomena such as aberration and total internal reflection in a broader class of geometries. Such techniques may prove to be useful from astrophysics, as light propagates through the intergalactic/galactic media; to the geometric optics of interfaces. 

The manuscript is structured as follows. In section \ref{sec.material} we recall the construction of a material manifold as the quotient space given by the orbits of the \emph{static} lab observer. Then, in section \ref{sec.stbton} we establish the relation between the unitary tangent bundle and the space of contact elements. In sections \ref{sec.euclid2d} - \ref{sec.lob}   we provide three specific examples for wavefront evolution in an optical medium by means of a contact transformation. The first two are optical media with Euclidian geometry in two and three dimensions. The results agree with Fermat's principle and recreate Snell's law of refraction. The third example, is a two dimensional optical medium endowed with an hyperbolic geometry.

\section{The material manifold}
\label{sec.material}

To model an optical medium will use a Riemannian manifold $(\mathcal{B}, g)$ submersed in a bi-metric  Lorentzian spacetime $(M,g_0,\tilde g)$, where $g_0$ is the background vacuum spacetime metric while $\tilde g$ represents the metric encoding the material properties of an optical medium. In this setting, the four velocity of the static observer in the \emph{lab frame},  $u\in TM$ satisfies \cite{LOPEZMONSALVO2020168270}
 	\beq
	g_0(u,u) = -1 \quad \text{and} \quad \tilde g(u,u)=-\frac{1}{n^2},
	\eeq
where $n>1$ is the refractive index of the medium.  In this way, the speed of light in the medium is clearly less than that in the vacuum background spacetime.  We use the vacuum metric $g_0$ to raise an lower indices by means of the musical isomorphisms
	\beq
	\label{eq.musical}
	g^\flat_0 : TM \longrightarrow T^*M \quad \text{and} \quad g^\sharp_0:T^*M\longrightarrow TM.
	\eeq

 Every non-spacelike vector field $w\in TM$  can be decomposed in terms of the \emph{optical metric} $\tilde g$ as
	\beq
	w = {\rm Hor}(w)+ {\rm Ver} (w) 
	\eeq
where ${\rm Hor}(w)$ and ${\rm Ver}(w)$ represent the transverse and parallel components of $w$ with respect to the lab frame $u$, respectively. Thus, every  null vector field $v$ such that
	\beq
	\label{eq.nullvert}
	{\rm Ver}(v) = \Omega u
	\eeq
for some non-vanishing function $\Omega$ on $M$, satisfies
	\begin{align}
	\tilde g(v,v) 	&= \tilde g\left({\rm Hor}(v),{\rm Hor}(v)\right) +  \tilde g\left({\rm Ver}(v),{\rm Ver}(v)\right)\nonumber\\
			&=  \tilde g\left({\rm Hor}(v),{\rm Hor}(v)\right) + \tilde g\left( \Omega u,  \Omega u \right)\nonumber\\
			&=\tilde g\left({\rm Hor}(v),{\rm Hor}(v)\right)  - \frac{\Omega^2}{n^2}=0.
	\end{align}
so that
	\beq
	\tilde g\left({\rm Hor}(v),{\rm Hor}(v)\right) = \frac{\Omega^2}{n^2}.
	\eeq

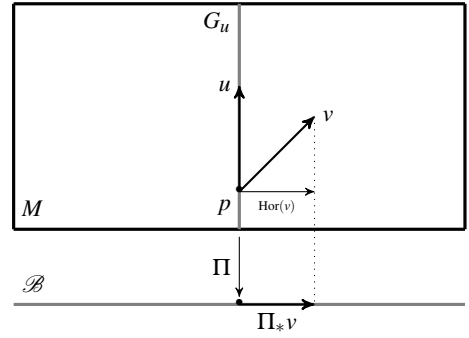
\begin{figure}
        \begin{center}
        \begin{tikzpicture}
		\draw[very thick] (0,0) -- (6,0);
		\draw[very thick] (0,0) -- (0,3);
		\draw[very thick] (0,3) -- (6,3);
		\draw[very thick] (6,3) -- (6,0);
		\draw[dotted] (4,1.5) -- (4,-1);
		\draw[->] (3,0.5) -- (4,0.5);
		\draw[very thick,color=gray] (0,-1) -- (6,-1);
		\draw[very thick,color=gray] (3,3) -- (3,0);
		\draw[->] (3,-0.1) -- (3,-0.9);
		\draw (0,0.5) node[anchor=north west] {$M$} ;
		\draw (0,-0.5) node[anchor=north west] {$\mathcal{B}$} ;
		\draw (3,3) node[anchor=north east] {$G_u$} ;
		\draw (3,0.5) node[anchor=north east] {$p$} ;
		\node at (3,0.5) {\textbullet};
		\node at (3,-1) {\textbullet};
		\draw[thick,->] (3,0.5) -- (4,1.5);
		\draw[thick,->] (3,0.5) -- (3,0.5+1.41);
		\draw (3,0.5+1.41) node[anchor=east] {$u$} ;
		\draw (4,1.5) node[anchor=west] {$v$} ;
		\draw[thick,->] (3,-1) -- (4,-1);
		\draw (3,-0.5) node[anchor= east] {$\Pi$} ;
		\draw (3.5,-1) node[anchor=north] {$\Pi_*v$} ;
		\draw (3.5,0.5) node[anchor=north] {\tiny ${\rm Hor}(v)$} ;
        \end{tikzpicture}
        \end{center}
        	\caption{Material manifold}
        	\label{fig.capcurve}
        \end{figure}

Now, let us consider the quotient manifold defined by the orbits of the lab frame four-velocity, namely
	\beq
	\mathcal{B}= M/G_u,
	\eeq
where $G_u$ represents the translation group associated to the flow of the vector field $u$.
One can equip this manifold with a Riemannian metric $g$ such that for every null vector field $v$ satisfying \eqref{eq.nullvert},
	\beq
	\label{eq.pushnull}
	g\left(\Pi_* v,\Pi_*v \right) =\frac{\Omega^2}{n^2},
	\eeq
where $\Pi: M \longrightarrow \B$ is a local trivialization. The pair $(\mathcal{B},g)$ is called a \emph{material manifold} Fig. \ref{fig.capcurve} (cf. definition of body manifold in \cite{carter1972foundations} and \cite{ehlers1973survey}). 

 The optical metric can be decomposed  in terms of the material manifold Riemannian metric and the observer's four velocity as
 	\beq
	\label{eq.opticmetric}
	\tilde g =  g \circ (\Pi_* \otimes \Pi_*) - \frac{1}{n^2} u^\flat \otimes u^\flat,  
	\eeq
where $u^\flat \equiv g^\flat_0(u)$ [cf equation \eqref{eq.musical}]. In the rest of the manuscript, we will consider the normalization factor $\Omega =1$.  Thus, expression \eqref{eq.opticmetric} corresponds  to the well known Gordon's optical metric \cite{schuster2018bespoke}.

In the following section, we will consider the unitary tangent bundle $ST\B$ together with ${PT^*}\mathcal{B}$ representing the  co-tangent bundle of $\mathcal{B}$ with the \emph{zero-section} $0_\mathcal{B}$ removed \cite{arnold1989mathematical}.
This allows us to represent the dual notions of light rays and wave fronts on the optical medium $(\B,g)$ \cite{VIArnold-LPDF,arnold1989mathematical,de1971gravitational}. The former one represents light rays as the projected null geodesic flow of $\tilde g$ while  the elements of ${PT^*}\mathcal{B}$ are the tangent planes to the wavefronts \cite{GeigesH-CHandCG}. Moreover, there is a one to one correspondence  between the elements of these two manifolds relating  a geodesic flow in the unitary tangent space to a contact transformation in the space of contact elements \cite{HGeiges-ABHCGandT}. Finally, since  contact transformations are  symmetries of  a contact distribution,  its  associated contact elements remain invariant when projected to the base manifold $\B$.


\section{Mathematical structures for rays and fronts.}
\label{sec.stbton}

In this section, we explicitly exhibit the dual nature of  light rays and wave fronts from the  contact geometric perspective. Let us consider the tangent bundle $T\B$ of an $m$-dimensional material manifold $(\B,g)$. Its associated  unitary tangent bundle is defined as 
	\beq
	\label{eq.stb}
	ST\B \equiv \{v\in T\B\ \vert\ {\bf g}(v,v) = 1 \},
	\eeq
where ${\bf g}$ is the \emph{bundle metric}  of $g$ \cite{jost2008riemannian}, (see Exercise 2 of Chapter 3, Section 4 in \cite{do1992riemannian}). 
It follows from the unitary constraint that 
	\beq
	{\rm dim}(ST\B) = 2m-1.
	\eeq
for $m$ the dimension of $\B$. Indeed, the defining property in \eqref{eq.stb} corresponds to the zero level set of a function on $T\B$, namely, ${\bf g}(v,v)-1=0$.

Similarly, the unitary co-tangent bundle $ST^*\B$ is defined as
	\beq
	ST^*\B = \{p \in T^*\B \vert {\bf g}^{-1}(p,p) = 1 \}.
	\eeq

Note that the restriction of  $ST^*\mathcal{B}$  to $PT^*\B$ corresponds to a \emph{ray-optical structure} -- as  defined by Perlick (see Definition 5.1.1 and Proposition 5.1.1 in  \cite{perlick2000ray}) -- that is,
	\beq
	\mathcal{N} = \{p \in PT^*\B \vert {\bf g}^{-1}(p,p) = 1\}.
	\eeq

The co-tangent bundle $T^*\B$  carries a natural symplectic structure, that is, a non-degenerate, closed 2-form $\omega$. Consider a vector field $L\in T(T^*\B)$ such that
	\beq
	\label{eq.liouvdef}
	\lie_L \omega= \omega,
	\eeq
where $\lie$ denotes the Lie derivative and $L$ is called a \emph{Liouville} vector field. One can define a 1-form $\lambda \in T^*(T^*\B)$ generating the symplectic structure in terms of the Liouville vector field as 
	\beq
	\label{liouville.coord}
	\lambda \equiv \dot\iota_L\omega.
	\eeq
It follows from Cartan's identity and definition \eq{liouville.coord} that
	\beq
	\lie_L \omega = \dot\iota_L \d \omega + \d \dot\iota_L \omega = - \d \dot\iota_L \omega = - \d \lambda = \omega.
	\eeq
In this way,  one can define a contact 1-form for $ST^*\B$, namely,
	\beq
	\eta = i^*(\lambda).
	\eeq
Here $i^*:T^*(T^*\B)\longrightarrow T\mathcal{N} $ is the pullback induced by the inclusion map $i:\mathcal{N}\longrightarrow T^*\B$. Let $\mathcal{D} = {\rm ker}(\eta)$  be the contact distribution generated by $\eta$, then $(\mathcal{N},\mathcal{D})$ is a \emph{contact manifold}, usually referred as the \emph{space of contact elements} or the \emph{contact bundle} of $\B$. The restricted bundle projection 
	\beq
	\pi\vert_\mathcal{N}:\mathcal{N}\subset T^*\B \longrightarrow \B
	\eeq  
maps the hyperplanes of the contact structure $\mathcal{D}$ to the contact elements of $\B$.  Note that the vector fields tangent to affinely parametrized geodesics on $(\B,g)$ are sections of $ST\B$. In this sense, we can establish a relation between $ST\B$ and $\mathcal{N}$ by demanding that
	\beq
	\label{eq.reebv}
	 g^{-1}\left(\Pi_* \circ \iota_*\right)(\xi)=\frac{1}{n}\left(\pi_* (v)\right) 
	\eeq
where  $v\in ST\B$, $n$ is a non-vanishing real number and $\xi$ is the \emph{Reeb} vector field  defined by
	\beq
	\label{eq.reeb}
	\dot\iota_\xi\eta = 1 \quad \text{and} \quad \dot\iota_\xi\d\eta = 0.
	\eeq
	
Thus, the required condition  \eqref{eq.reebv} expresses that the geodesics of $(\B,g)$ are transverse to $\mathcal{N}$ and that $\xi$ is the spatial part of a null vector field in $TM$ [cf. equation \eqref{eq.pushnull}]. Indeed,
	\beq
	{\bf g}(\xi,\xi) = g(\xi\vert_x,\xi\vert_x) = \frac{1}{n^2} \quad \text{for each} \quad x \in \B.
	\eeq
This, however, does not imply that the light rays should be metric orthogonal to $\mathcal{N}$. The orthogonality condition implies the existence of an \emph{almost contact structure} 
	\beq
	\phi:T\mathcal{N} \longrightarrow T\mathcal{N},
	\eeq
where
	\beq
	\label{eq.acs}
	\phi^2 = -\mathbbm{1} + \eta \otimes \xi \quad \text{and} \quad \eta \circ \phi = 0,
 	\eeq
	 such that $(\mathcal{N},\eta, {\bf g}\vert_\mathcal{N})$ is a contact metric manifold. That is, the restricted bundle metric ${\bf g}\vert_\mathcal{N} = i^*({\bf g})$ can be written as
	\beq
	{\bf g}\vert_\mathcal{N} = \eta \otimes \eta + \d \eta \circ (\phi \otimes \mathbbm{1}).
	\eeq
Indeed, 
	\beq
	{\bf g}\vert_\mathcal{N}(\mathcal{D},\xi) = \eta(\mathcal{D})\eta(\xi) + \d \eta(\phi\mathcal{D},\xi) =0,
	\eeq
where the last equality follows from the transversality condition \eqref{eq.reeb} and the definition $\phi$, equation \eqref{eq.acs} where $\phi \mathcal{D} =\mathcal{D}$. In the next section we will see that the presence of boundaries (interfaces between different media) breaks the orthogonality condition.

Now, let us recall the definition of a \emph{wavefront} centered at the point $b \in \B$ as the surface 
	\beq
	F_b(t)=\{b_i \in \B \,\vert \gamma(0)=b, \gamma(t)=b_i\} 
	\eeq
where $\gamma$ is a geodesic  on $(\B,g)$ (see \cite{HGeiges-ABHCGandT}). Observe that for a point $b_i \in F_b(t)$ the contact element $\pi(\mathcal{D}_{b_i} )$ is tangent to $F_b(t)$. The bundle projector $\pi$ projects Legendrian submanifolds of $\mathcal{N}$ to wavefronts in $\B$.

\begin{figure}
\begin{center}
\includegraphics[width=0.85\columnwidth]{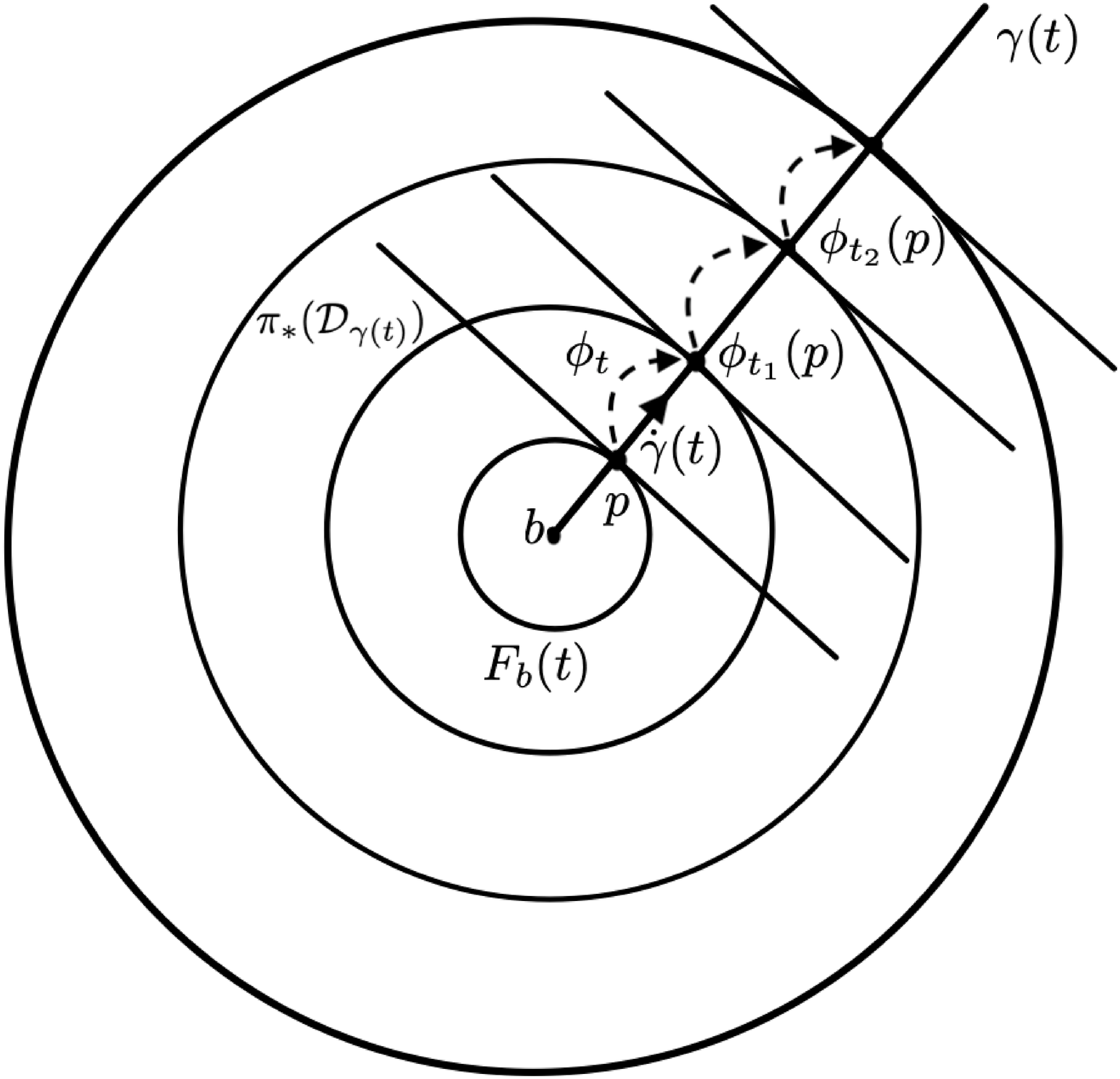}
\caption{Representation of the contact transformation overlapped with the wavefronts and the contact elements, for which $b_i=\gamma(t_i)$. If the medium is simply connected (there are no interfaces), then wavefronts are perpendicular to light rays fulfilling equation \eq{light.perp.wave}. Nevertheless, it always satisfies that for $p \in \gamma$ then $\phi_t(p) \in \gamma$.}

\label{contact.trans}
\end{center}
\end{figure}

Consider a geodesic vector flow on $ST\B$ given by $\gamma(t) \in \B$ and $\dot \gamma(t) \in ST_{\gamma(t)}\B$. There is a unique contact element such that
	\beq
	\label{light.perp.wave}
	 {\bf g}|_{\mathcal N} \left(\mathcal D_{b_i},\xi{|_{b_i}} \right)=0.
	\eeq
The Reeb flow induces a strict contact transformation
	\beq
	\lie _{\xi_\lambda}\lambda=0.
	\eeq
Thus, by the duality of the Reeb's and the geodesic flow (see Theorem 1.5.2 in \cite{HGeiges-ContTopo}), there exists a 1-parameter family of contact transformations $\phi_t: \mathcal{N} \to \mathcal{N}$, such that for every $p \in \mathcal{N}$ where $p=g^\flat\left(\dot \gamma (0)\right)$, 
	\beq
	\phi_t(p)= g^\flat\left(\dot \gamma (t)\right)
	\eeq
is satisfied. 
Therefore, the geodesic flow can be evolved in terms of the 1-parameter contact transformation $\phi_t$ in $\mathcal{N}$ (see Fig. \ref{contact.trans}) induced by the Reeb vector field. In the next sections, we will make some explicit examples for optical media represented by different geometries such as Euclidean and hyperbolic ones. We will also explore some different dimensions and interfaces between media. 

. 
 		
 		
\section{Optical medium in Euclidean $\mathbb{R}^2$}
\label{sec.euclid2d}

Let us consider an optical medium with electric permittivity $\varepsilon$ and magnetic permeability $\mu$ so that its refractive index is given by $n^2=\varepsilon \mu$. The Euclidean optical metric is written as
	\beq
	\tilde g=\sum_{i=1}^2 \d x^i \otimes \d x^i -\frac{1}{n^2}\d t \otimes \d t.
	\eeq
Restricting  our attention to  the  purely spatial part of the metric by means of a conformal projection into a 2-dimensional manifold, the material manifold metric becomes
	\beq
	g=n^2\, \sum_{i=1}^2 \d x^i \otimes \d x^i. 
	\eeq

The unitary tangent bundle $ST\re^2$ is expressed in local coordinates $\{x, y, p_x, p_y\}$ where the standard symplectic 2-form is written as 
	\beq
	\omega= \sum_{i=1}^2 \d x_i \wedge \d p^i.
	\eeq
	  The unitary condition \eq{eq.stb} leads to
	\beq
	\label{euclid.norm}
	p_y=\sqrt{n^2-p_x^2}.
	\eeq

Consider the  transformation $\phi_{stb}: T\re^2 \to ST\re^2$ defined by
	\beq
	 \label{stb.trans}
	\left[x=x,\;  y=y, \; p_x=-\sin{\theta},\; p_y=\cos{\theta}\right]
	\eeq
to change into polar coordinates $\{x,y,\theta\}$. In this coordinates, the Liouville 1-form \eq{liouville.coord} becomes



	\beq
	\lambda_{stb}=-\sin{\theta}\,\d x+\sqrt{\cos^2{\theta}+n^2-1}\, \d y,
	\eeq
and the Reeb vector field associated to $\lambda_{stb}$  is
	\beq
	\xi_{stb}=\frac{1}{n^2}\left(-\sin{\theta}\,\prt{x}+\sqrt{\cos^2{\theta}+n^2-1}\, \prt{y}\right).
	\eeq	

 A 1-parameter family of strict contact transformations induced by the Reeb's flow is given by
	\beq
	\label{cont.trans}
	\phi_t=\left[x=x-\frac{t\, \sin{\theta}}{n^2},y=y+\frac{t\, \cos{\theta}}{n^2}, \theta=\theta \right].
	\eeq

The flow of the Reeb vector field, as proved by Geiges \cite{HGeiges-ContTopo}, results dual to the geodesic flow. Observe that \eq{cont.trans} is indeed a strict contact transformation, as $\phi_t^*(\lambda)=\lambda$. In terms of \eq{cont.trans}, the contact distribution transforms into 
 	\beq
	\label{contact.dist}
	\chi={\rm span}\left\{\frac{1}{n^2} \sqrt{\cos^2{\theta}+n^2-1}\; \prt{x}-\sin{\theta}\,\prt{y},\; \prt{\theta} \right\} .
	\eeq

The bundle projection of the contact distribution $\pi(\chi)$ are the contact elements of $\re^2$. In this case, this corresponds to the first vector of \eq{contact.dist}. For a point $q\in \re^2$ and a fixed $\theta_0$ there is a unique geodesic   passing through $q$ and the direction of its tangent vector is precisely $\theta_0$. Thus, the contact element is  the positive normal line  generated by the vector $\pi(\chi)$. \\

\begin{figure}
\begin{center}
\includegraphics[width=0.85\columnwidth]{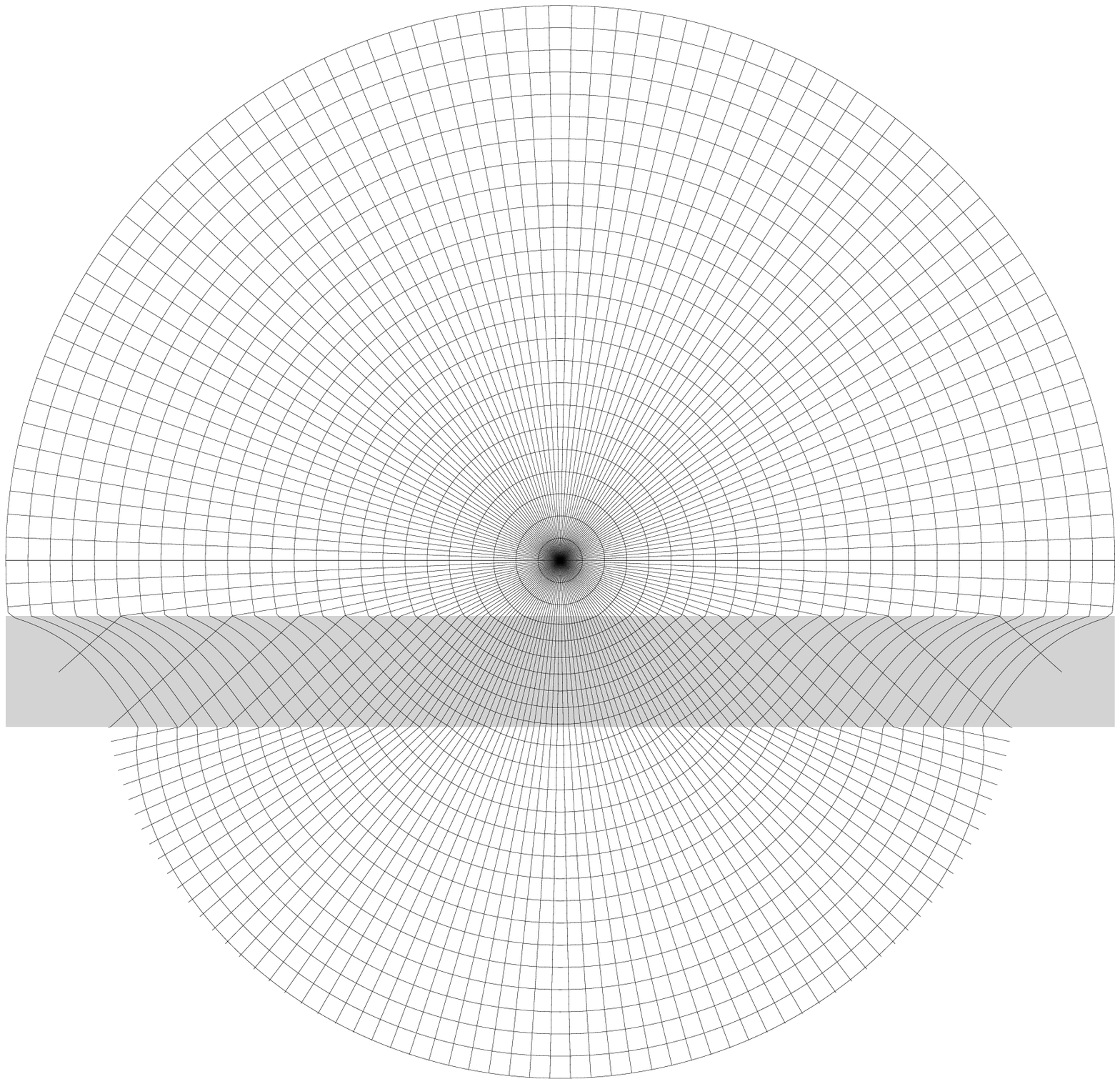}
\caption{Light rays and wavefronts emitted by a source of light in air ($n=1$) passing through a layer of water ($n=1.33$) represented by the gray layer. The light rays refract as predicted by Snell's law. Wavefronts are deformed as they are slowed down by the interface. After returning to air, wavefronts do not return to its original shape.}
\label{snell.law}
\end{center}
\end{figure}

\begin{figure}
\begin{center}
\includegraphics[width=0.85\columnwidth]{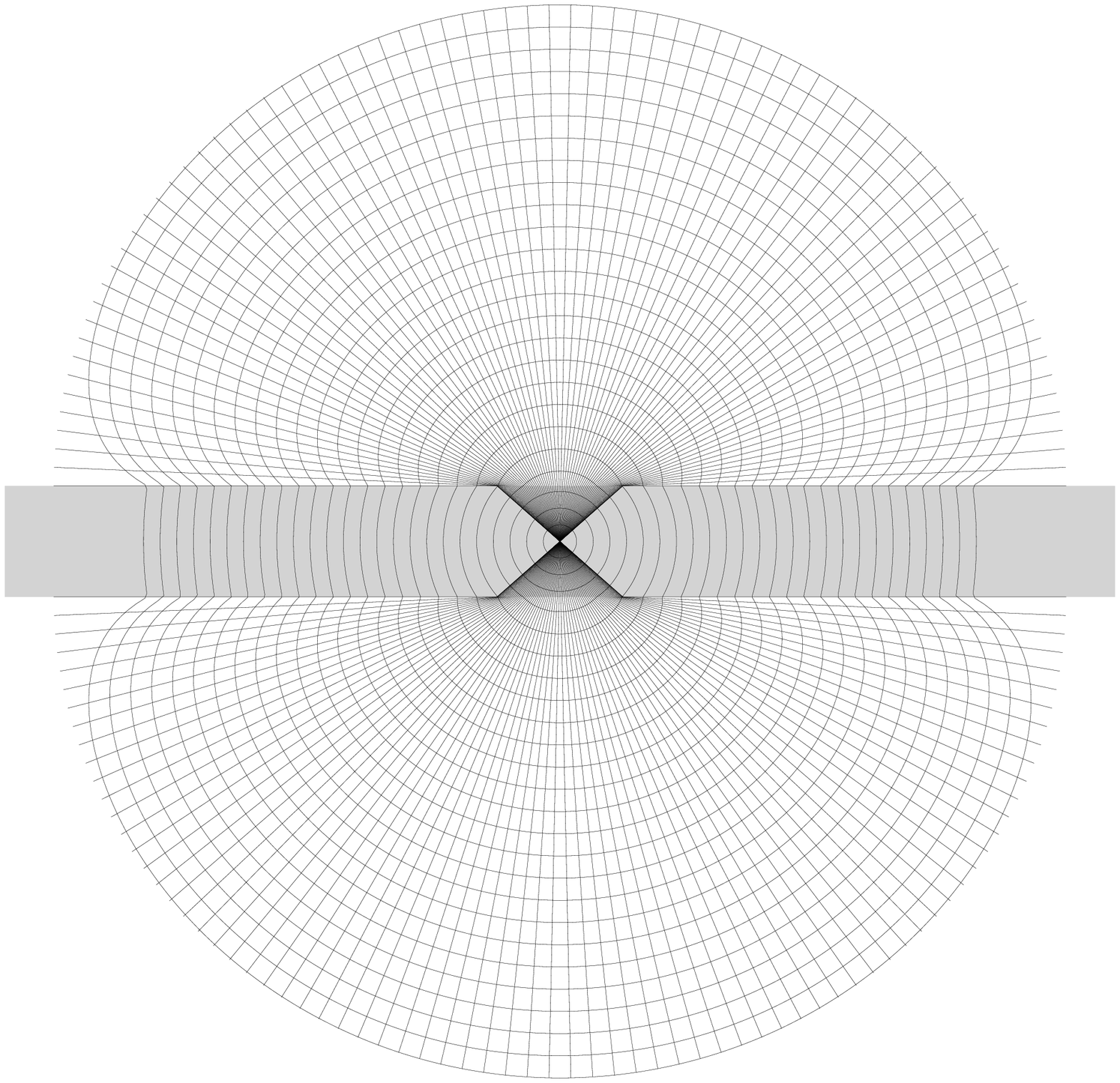}
\caption{Light rays and wavefronts emitted by a source of light in water ($n=1.33$) represented by the gray layer. When light rays refract after passing through the interface with aire ($n=1$), there is a critical angle $\theta_c$ for which light rays do not come out from the water. In the figure, only the refracted light rays are shown. Wavefronts, deform as they pass through the interfase. After leaving the first medium, wavefronts  are not spherical anymore. Some aberrations can be observed as light rays refracted almost parallel to the surface of the interface do not cross perpendicular the wavefronts.}
\label{inner.total}
\end{center}
\end{figure}

The geodesic flow is now written in terms of the 1-parameter family of strict contact transformations \eq{cont.trans}. Observe that for vacuum ($\varepsilon_0 \mu_0=1$) the contact transformation will induce a flow given by a family of straight lines with slope $m=-(\tan{\theta_0})^{-1}$, recalling that $\theta_0$ is the angle of the contact element positively ortogonal to the geodesic. As the geodesic curve is the trajectory of a light beam, when it pases to a different medium ($\varepsilon \mu\neq1$), the light will be refracted with $n=\sqrt{\varepsilon \mu}$ as expected by Snell's law. Furthermore,  we can reconstruct the wavefronts form the contact elements without solving directly the wave equation. In Fig.\ref{snell.law}, we can observe how the wavefronts are deformed and slowed down when passing through a different medium. In Fig. \ref{inner.total} the source of light is in a medium with a larger refractive index. Note that when the light passes through the interface, \emph{total internal reflection} can be observed in the light rays. 


\section{Optical medium in Euclidean $\mathbb{R}^3$} 

In the same maner as before, let us consider an optical medium in $\re^3$ with body metric
	\beq
	g=n^2 \sum_{i=1}^3 \d x^i \otimes \d x^i .
	\eeq
Again, we propose the coordinate transformation to the unitary tangent bundle $\phi_{stb}: T\re^3 \to ST\re^3$
	\bea
	\label{stb.trans.r3}
	&\left[x=x,\;  y=y, \; z=z,\; p_x=\sin{\theta}\cos{\varphi},\right.\nonumber\\ 
	 & \left. \; p_y=\sin{\theta}\sin{\varphi}, \; p_z=\cos{\theta}\right]
	\end{align}
to get spherical coordinates $\{x, y, z, \theta, \varphi\}$. The Liouville 1-form transforms to
	\beq
 	\lambda_{stb}=\cos{\varphi}\sin{\theta}\,\d x+ \sin{\varphi}\sin{\theta}\,\d y+\sqrt{\cos^2{\theta}+n^2-1}\, \d z,
	\eeq
for which the Reeb vector field is
	\bea
	\label{r.flow.r3}
	\xi_{stb}=\frac{1}{n^2}&\left(\cos{\varphi}\sin{\theta}\,\prt{x}+ \sin{\varphi}\sin{\theta}\,\prt{y} \right. \\ \nonumber
	 & \left. +\sqrt{\cos^2{\theta}+n^2-1}\, \prt{z}\right)
	\end{align}
while the contact distribution is expressed as
	\bea 
	\chi={\rm span}&\left\{ -\sin{\theta}\sin{\varphi}\; \prt{x}+\cos{\varphi}\sin{\theta} \; \prt{y}, \; 
	  \right. \\ \nonumber
	 & \left.  -\sqrt{\cos^2{\theta}+n^2-1}\; \prt{x} + \cos{\varphi}\sin{\theta}\; \prt{z}, \;  \prt{\theta}, \; \prt{\varphi} \right\}.
	\end{align}

We find the 1-parameter family of contact transformations induced by the Reeb vector field \eq{r.flow.r3} 
	\bea
	\label{cont.trans.r3}
	\phi_t=&\left[x=x-\frac{t\,\cos{\varphi} \sin{\theta}}{n^2},\;y=y+\frac{t\, \sin{\varphi} \sin{\theta}}{n^2}, \right.\\ \nonumber
	&\left. z=z+\frac{t\,\sqrt{\cos^2{\theta}+n^2-1}}{n^2} ,\;  \theta=\theta,\; \varphi=\varphi\right].
	\end{align}

As expected, the projection of the contact distribution to $\re^3$ is a set of 2-dimensional planes, which are the contact elements of $\re^3$. Each plane is tangent to a wavefront surface sphere. 

\section{Optical medium with hyperbolic geometry on $\mathbb{R}^2$ half plane}
\label{sec.lob}

We consider now a 2-dimensional optical medium endowed with a hyperbolic geometry represented by the upper half plane model $\mathbb{H}^2=\{z=x+iy,\,  y>0\}$ with the usual metric
	\beq
	\tilde g=\frac{1}{y^2}\left(\sum_{i=1}^2 \d x^i \otimes \d x^i-\frac{1}{n^2} \d t \otimes \d t\right).
	\eeq
As before, we will work with the body metric given by
	\beq
	g=\left(\frac{n}{y}\right)^2\,\sum_{i=1}^2 \d x^i \otimes \d x^i.
	\eeq

The unitary tangent bundle of the hyperbolic space is naturally identified with $PSL(2,\re)$, which is different from $ST\re^n$ used in the previous examples. Nevertheless, our construction is sufficiently general  to work in either space, as $PSL(2,\re)$ can also be endowed with coordinates $\{x,y,\varphi\}$, for $(x,y)$ cartesian coordinates and $\varphi$ an angular coordinate \cite{Bolsinov2019ChaosAI}. In terms of these coordinates, the Liouville 1-form is 

 	\beq
 	\lambda_{stb}=\frac{n}{y}\left(- \cos{\varphi}\; \d x+ \sin{\varphi}\; \d y\right),
	\eeq
The associated Reeb vector field is 
	\beq
	\label{reeb.stb.hyperbolic}
	\xi_{stb}=\frac{1}{n}\left(-y \cos{\varphi}\; \prt{x}+ y \sin{\varphi}\; \prt{y}- \cos{\varphi} \; \prt{\varphi}\right),
	\eeq
and the contact distribution
	\beq
	\chi={\rm span}\left\{ \frac{y}{n}\left(\sin{\varphi}\; \prt{x} + \cos{\varphi}\; \prt{y}\right),\; \prt{\varphi}\right\}.
	\eeq 

Once again, we find the 1-parameter family  of contact transformations in $\text{P}T^*\mathcal{B}$ induced by the Reeb's flow with the natural identification with $ST\mathbb{H}^2$.  
	\bea
	\label{lob.contact.trans}
	\phi_t=&\left[x=\frac{(x\sin{\varphi}+y \cos{\varphi}-x)e^{\frac{2t}{n}}-x\sin{\varphi}-y\cos{\varphi}-x}{(\sin{\varphi}-1)e^{\frac{2t}{n}}-\sin{\varphi}-1}, \right.\\ \nonumber
	& \left. y=-\frac{2y e^{\frac{t}{n}}}{(\sin{\varphi}-1)e^{\frac{2t}{n}}-\sin{\varphi}-1}\right. ,\\ \nonumber
	&\left.\;\varphi= \arctan{\left(-\frac{(-\sin{\varphi}+1)e^{\frac{2t}{n}}-\sin{\varphi}-1}{2e^{\frac{t}{n}}\cos{\varphi}}\right)}\right]
	\end{align}

\begin{figure}
\begin{center}
\includegraphics[width=0.8\columnwidth]{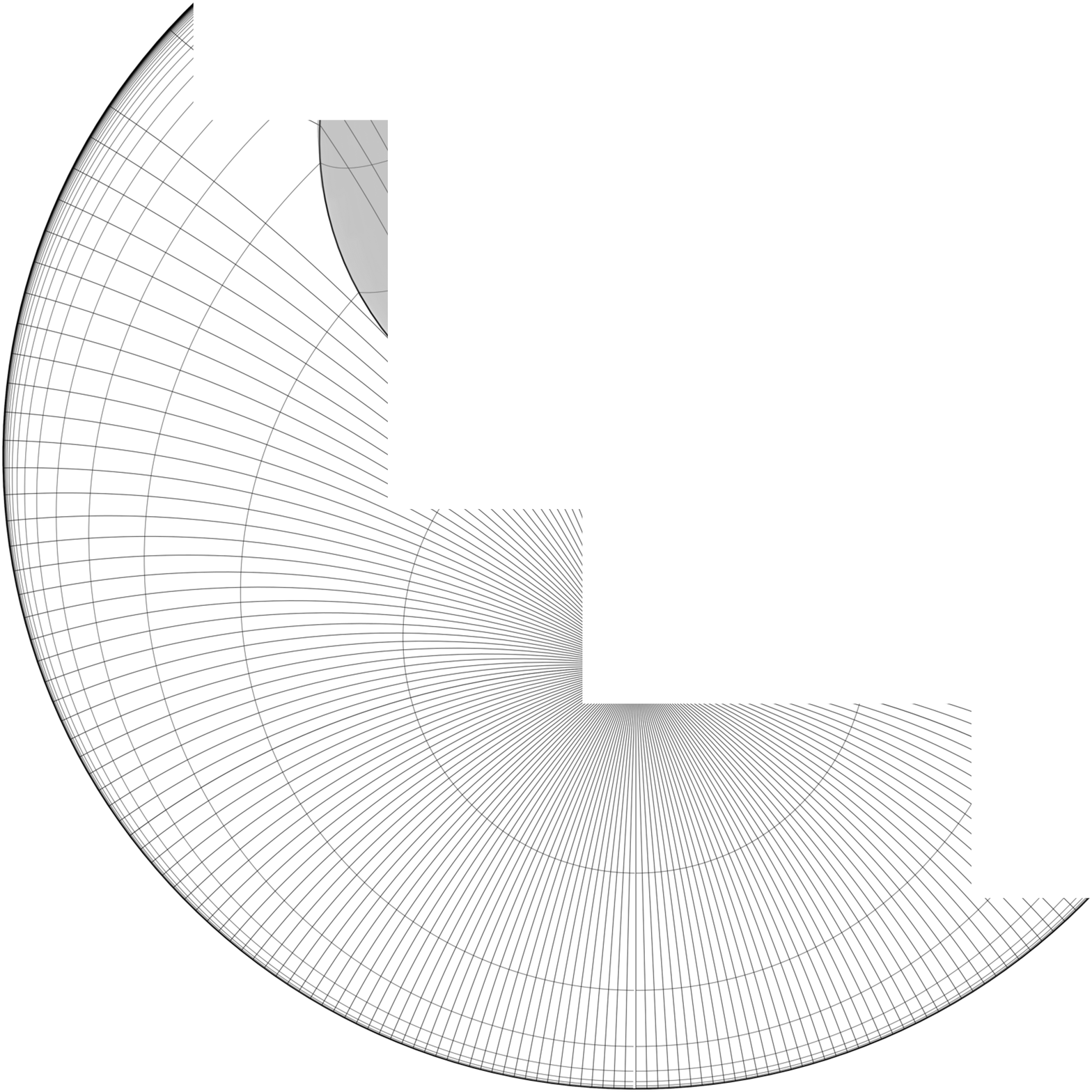}
\caption{Light rays and wavefronts emitted by a source of light in air ($n=1$) refracting through a horizontal layer of water ($n=1.33$) in the upper half plane mapped to the Poincar\'e disc.}
\label{hyp.ref.hor}
\end{center}
\end{figure}

\begin{figure}
\begin{center}
\includegraphics[width=0.8\columnwidth]{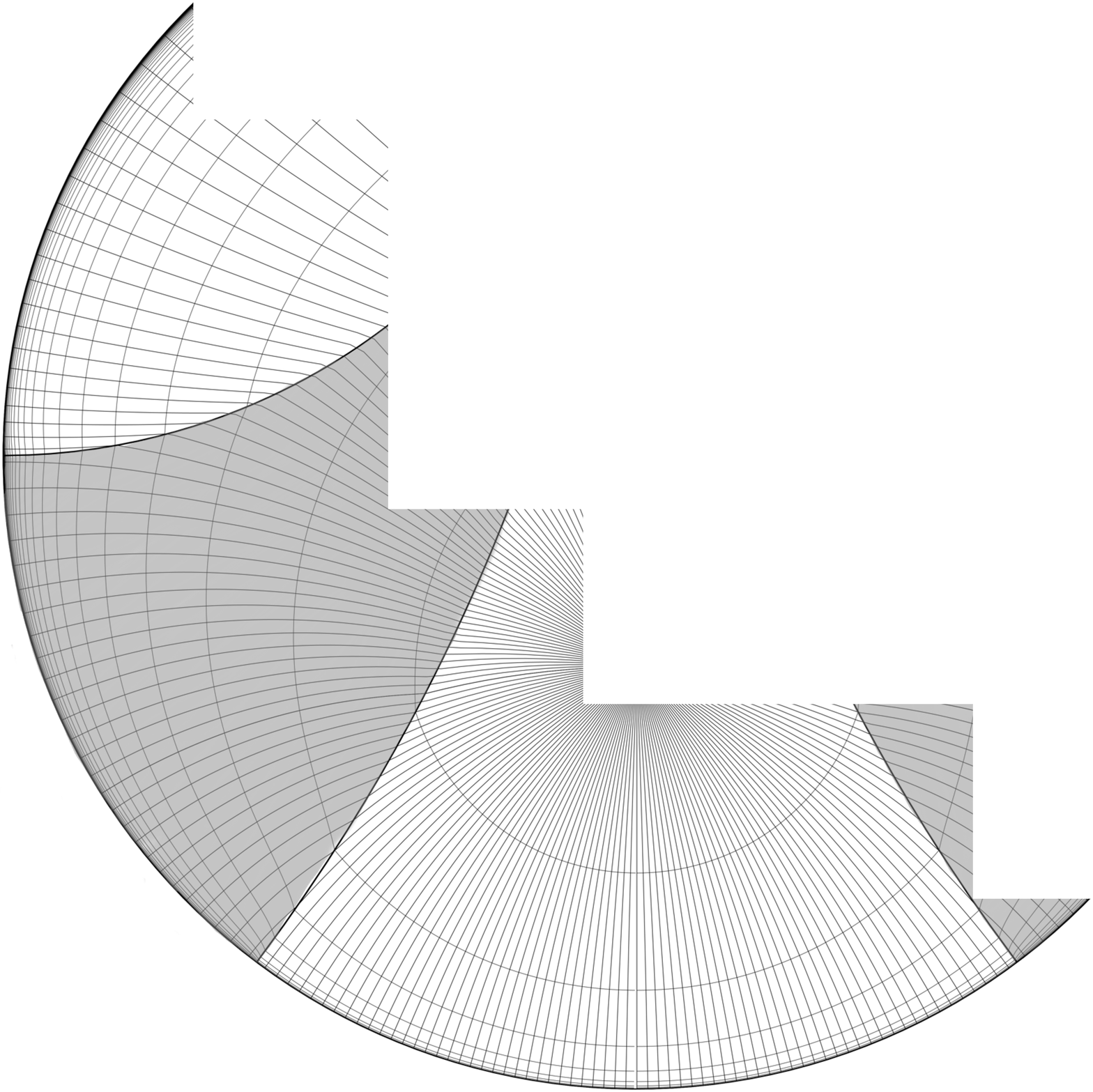}
\caption{Light rays and wavefronts emitted by a source of light in air ($n=1$) refracting through a vertical layer of water ($n=1.33$) in the upper half plane mapped to the Poincar\'e disc.}
\label{hyp.ref.vert}
\end{center}
\end{figure}

The mapping (\ref{lob.contact.trans}) is indeed a strict contact transformation as $\phi^* _t(\lambda)=\lambda$. 
As expected, the projection of the Reeb's flow which induced \eq{lob.contact.trans} are semicircles of radius 
	\beq
	r=\frac{y_0}{\cos{\varphi_0}}
	\eeq
and center
	\beq
	\left(\frac{x_0\,\cos{\varphi_0}-y_0\,\sin{\varphi_0}}{\cos{\varphi_0}}, 0\right),
	\eeq
which are precisely the geodesic curves in $\mathbb{H}^2$. 

For interfaces between media with different refractive index, we solve \eq{reeb.stb.hyperbolic} using numerical methods. The solution observed in Fig \ref{hyp.ref.hor} and \ref{hyp.ref.vert} represents the refraction of the light rays mapped to the Poincar\'e disc, the first one represents a horizontal layer of a different refractive index as interface, while the second represents a vertical interface. 

\begin{figure}
\begin{center}
\includegraphics[width=0.85\columnwidth]{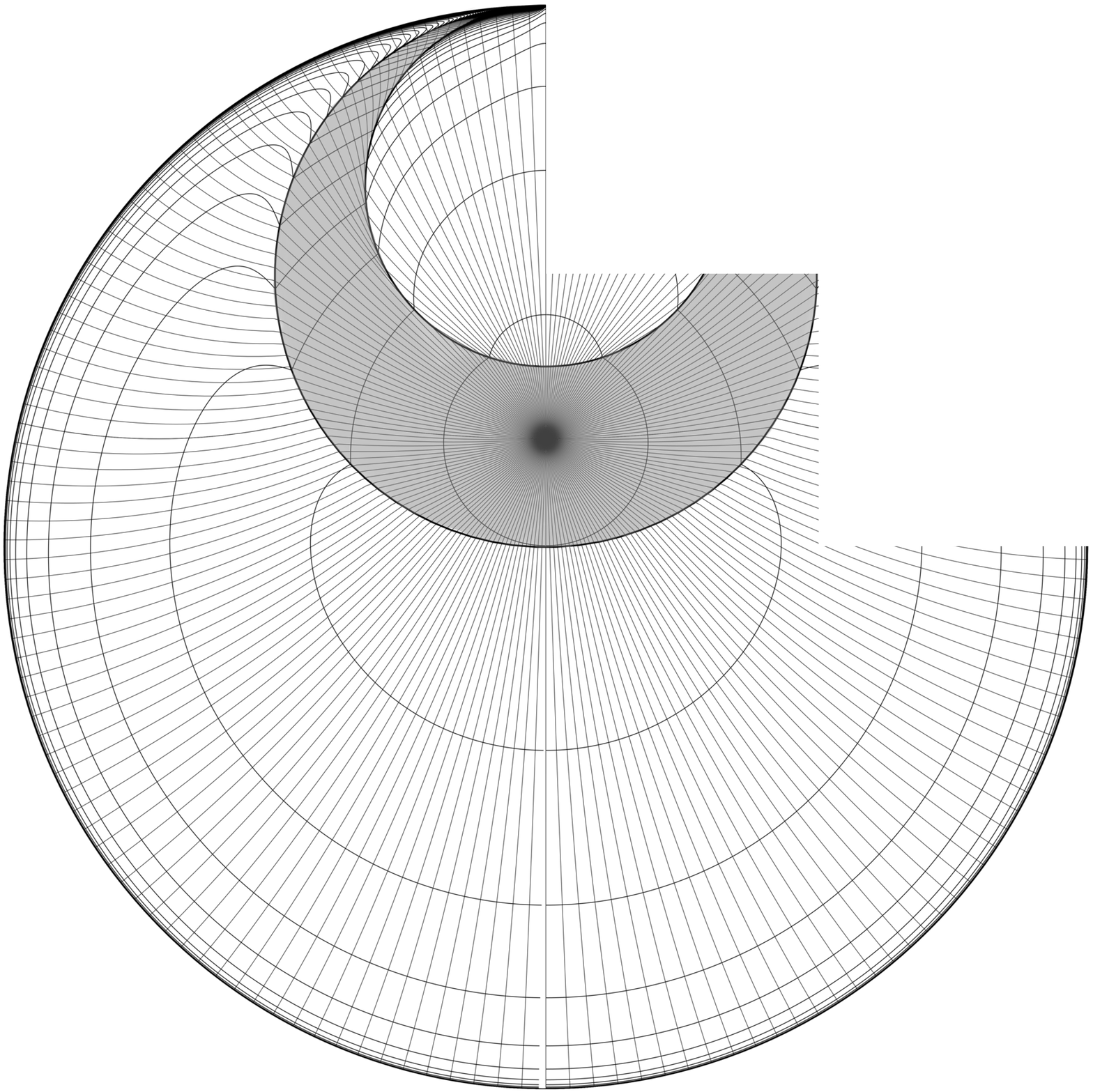}
\caption{Light rays and wavefronts emitted by a source of light which is in water ($n=1.33$) refracting through a horizontal layer to air ($n=1$) in the upper half plane mapped to the Poincar\'e disc. Observe that no total internal reflection is observed. As the light rays are closer to the circule boundary of the disc, some of them intersects more than one time the same wavefront, which means that light rays in different time intersects the same hypersurface of constant time. This type of aberrations are due to the geometry and can be interpreted as achronal surfaces.}
\label{hyp.tot.ref}
\end{center}
\end{figure}

Finally, we use the same numerical methods to obtain the waves and fronts of a  light source in the medium with larger refractive index. The result, shown in Fig \ref{hyp.tot.ref}, presents no total internal reflection. 

In the gravitational context, the anti-deSitter model has been closely related with the hyperbolic geometry. It is known that AdS  admits  closed timelike curves and achronal surfaces can be observed. In Fig \ref{hyp.tot.ref} we can observe some of this geometric properties in light rays which are closer to the infinity circle boundary of the disc. Where one light ray can intersect more than once the same wavefront. 

\section{Closing remarks}

In this work we provide a direct application of metric contact geometry to model light propagation in materials whose refractive properties are modelled by metric tensors. Albeit the duality between Fermat and Huygens' principles has been extensively explored, the present study is, to the best of our knowledge, the first to use the richer metric structure of contact geometry. Moreover, this allowed us to explore light propagation through interfaces for both, rays and fronts. These numerical explorations showed that the `well stitched' geometric structure, responsible for the orthogonality between rays and fronts, is broken when one considers interfaces or boundaries. The formal aspects of including boundaries in metric contact manifolds will be studied elsewhere.

The spaces  $ST\B$ and $\mathcal{N}$  are the mathematical structures required to express the propagation of light through the two key principles of Fermat and Huygens. On the one hand,  $ST\B$  is the spatial projection of the null cone where light rays correspond to null curves on $(M,g_0,\tilde g)$, allowing a spacetime formulation of Fermat's principle. On the other hand, $\mathcal{N}$ serves us to define the wave fronts  tangent to the contact elements on $\B$. The evolution of light rays in $ST\B$ is given by the flow of the Reeb vector field while, in $\mathcal{N}$, it is given by its associated contact transformation. In this sense, the entire nature of light propagation in non-dispersive media is encoded in the contact bundle of $(\B,g)$ together with its symmetries.

Decomposing the optical metric in terms of the observer's velocity and the Riemannian metric of the material manifold, together with the contact structure associated with the ray-optical structure $\mathcal{N}$ allows one to relate it with the unitary tangent bundle of $\B$, demanding that the spatial projection of the Reeb vector field be proportional to the projection of a null vector field of the background spacetime. This relation implies that the Reeb's flow represents indeed the light trajectory on the optical medium $\B$. 


The Reeb flow allows one to reconstruct the trajectories and wavefronts of light while propagating through an optical medium. Using this result, we explored two and three dimensional Euclidean optical media and a two dimensional hyperbolic medium. In the case of a 2-dimensional Euclidean medium, it was possible to recreate Snell's law of refraction together with the total internal reflection phenomenon. In the case of the hyperbolic medium, we obtained the refraction patterns from vertical and horizontal interfaces mapped to the Poincar\'e disc. No total internal reflection was observed when the light source was placed in a medium with a larger refractive index. Nevertheless, we saw that light rays can intersect the same wavefront more than once. This is a surprising effect which deserves further exploration. 


\section*{Acknowledgments}
DGP thanks Dr. Alessandro Bravetti for enlightenment discussions and observations. DGP was funded by a CONACYT Scholarship with CVU 425313. 


\begin{thebibliography}{10}

\bibitem{gordon1923lichtfortpflanzung}
W.~Gordon, ``Zur lichtfortpflanzung nach der relativit{\"a}tstheorie,'' {\em
  Annalen der Physik}, vol.~377, no.~22, pp.~421--456, 1923.

\bibitem{de1971gravitational}
F.~de~Felice, ``On the gravitational field acting as an optical medium,'' {\em
  General Relativity and Gravitation}, vol.~2, no.~4, pp.~347--357, 1971.

\bibitem{ehlers2012republication}
J.~Ehlers, F.~A. Pirani, and A.~Schild, ``Republication of: The geometry of
  free fall and light propagation,'' {\em General Relativity and Gravitation},
  vol.~44, no.~6, pp.~1587--1609, 2012.

\bibitem{pendry2006controlling}
J.~B. Pendry, D.~Schurig, and D.~R. Smith, ``Controlling electromagnetic
  fields,'' {\em science}, vol.~312, no.~5781, pp.~1780--1782, 2006.

\bibitem{leonhardt2006general}
U.~Leonhardt and T.~G. Philbin, ``General relativity in electrical
  engineering,'' {\em New Journal of Physics}, vol.~8, no.~10, p.~247, 2006.

\bibitem{chen2010transformation}
H.~Chen, C.~T. Chan, and P.~Sheng, ``Transformation optics and metamaterials,''
  {\em Nature materials}, vol.~9, no.~5, pp.~387--396, 2010.

\bibitem{LOPEZMONSALVO2020168270}
C.~Lopez-Monsalvo, D.~Garcia-Pelaez, A.~Rubio-Ponce, and R.~Escarela-Perez,
  ``The geometry of induced electromagnetic fields in moving media,'' {\em
  Annals of Physics}, vol.~420, p.~168270, 2020.

\bibitem{PhysRevA.102.023528}
H.~Chen, S.~Tao, J.~B\ifmmode~\check{e}\else \v{e}\fi{}l\'{\i}n, J.~Courtial,
  and R.-X. Miao, ``Transformation cosmology,'' {\em Phys. Rev. A}, vol.~102,
  p.~023528, Aug 2020.

\bibitem{schuster2019electromagnetic}
S.~Schuster and M.~Visser, ``Electromagnetic analogue space-times, analytically
  and algebraically,'' {\em Classical and Quantum Gravity}, 2019.

\bibitem{faccio2013analogue}
D.~Faccio, F.~Belgiorno, S.~Cacciatori, V.~Gorini, S.~Liberati, and
  U.~Moschella, {\em Analogue gravity phenomenology: analogue spacetimes and
  horizons, from theory to experiment}, vol.~870.
\newblock Springer, 2013.

\bibitem{schuster2018bespoke}
S.~Schuster and M.~Visser, ``Bespoke analogue space-times: Meta-material
  mimics,'' {\em General Relativity and Gravitation}, vol.~50, no.~6,
  pp.~1--21, 2018.

\bibitem{GeorgantzisGarcia:20}
D.~G. Garcia, G.~J. Chaplain, J.~B\v{e}l\'{i}n, T.~Tyc, C.~Englert, and
  J.~Courtial, ``Optical triangulations of curved spaces,'' {\em Optica},
  vol.~7, pp.~142--147, Feb 2020.

\bibitem{perlick2000ray}
V.~Perlick, {\em Ray optics, Fermat’s principle, and applications to general
  relativity}, vol.~61.
\newblock Springer Science \& Business Media, 2000.

\bibitem{berest1994huygens}
Y.~Y. Berest and A.~P. Veselov, ``Huygens' principle and integrability,'' {\em
  Russian Mathematical Surveys}, vol.~49, no.~6, pp.~5--77, 1994.

\bibitem{harte2013tails}
A.~I. Harte, ``Tails of plane wave spacetimes: Wave-wave scattering in general
  relativity,'' {\em Physical Review D}, vol.~88, no.~8, p.~084059, 2013.

\bibitem{mclenaghan1969explicit}
R.~McLenaghan, ``An explicit determination of the empty space-times on which
  the wave equation satisfies huygens' principle,'' in {\em Mathematical
  Proceedings of the Cambridge Philosophical Society}, vol.~65, pp.~139--155,
  Cambridge University Press, 1969.

\bibitem{noonan1995huygens}
T.~W. Noonan, ``Huygens' principle in conformally flat spacetimes,'' {\em
  Classical and Quantum Gravity}, vol.~12, no.~4, p.~1087, 1995.

\bibitem{kholodenko2013applications}
A.~L. Kholodenko, {\em Applications of contact geometry and topology in
  physics}.
\newblock World Scientific, 2013.

\bibitem{garcia2014infinitesimal}
D.~Garc{\'\i}a-Pel{\'a}ez and C.~Lopez-Monsalvo, ``Infinitesimal legendre
  symmetry in the geometrothermodynamics programme,'' {\em Journal of
  Mathematical Physics}, vol.~55, no.~8, p.~083515, 2014.

\bibitem{bravetti2015contact}
A.~Bravetti, C.~Lopez-Monsalvo, and F.~Nettel, ``Contact symmetries and
  hamiltonian thermodynamics,'' {\em Annals of Physics}, vol.~361,
  pp.~377--400, 2015.

\bibitem{bravetti2017contact}
A.~Bravetti, H.~Cruz, and D.~Tapias, ``Contact hamiltonian mechanics,'' {\em
  Annals of Physics}, vol.~376, pp.~17--39, 2017.

\bibitem{lopez2021contact}
C.~Lopez-Monsalvo, F.~Nettel, V.~Pineda-Reyes, and L.~Escamilla-Herrera,
  ``Contact polarizations and associated metrics in geometric thermodynamics,''
  {\em Journal of Physics A: Mathematical and Theoretical}, vol.~54, no.~10,
  p.~105202, 2021.

\bibitem{flores2021contact}
D.~Flores-Alfonso, C.~S. Lopez-Monsalvo, and M.~Maceda, ``Contact geometry in
  superconductors and new massive gravity,'' {\em Physics Letters B}, vol.~815,
  p.~136143, 2021.

\bibitem{PhysRevLett.127.061102}
D.~Flores-Alfonso, C.~S. Lopez-Monsalvo, and M.~Maceda, ``Thurston geometries
  in three-dimensional new massive gravity,'' {\em Phys. Rev. Lett.}, vol.~127,
  p.~061102, Aug 2021.

\bibitem{HGeiges-ABHCGandT}
H.~Geiges, ``A brief history of contact geometry and topology,'' {\em
  Expositiones Mathmaticae}, vol.~19, pp.~25--53, 2001.

\bibitem{carter1972foundations}
B.~Carter and H.~Quintana, ``Foundations of general relativistic high-pressure
  elasticity theory,'' {\em Proceedings of the Royal Society of London. A.
  Mathematical and Physical Sciences}, vol.~331, no.~1584, pp.~57--83, 1972.

\bibitem{ehlers1973survey}
J.~Ehlers, ``Survey of general relativity theory,'' in {\em Relativity,
  astrophysics and cosmology}, pp.~1--125, Springer, 1973.

\bibitem{arnold1989mathematical}
V.~I. Arnold, {\em Mathematical methods of classical mechanics}, vol.~60.
\newblock Springer, 1989.

\bibitem{VIArnold-LPDF}
V.~I. Arnold, {\em Lectures on Partial Differential Equations}.
\newblock Springer, 2014.

\bibitem{GeigesH-CHandCG}
H.~Geiges, ``Christiaan huygens and contact geometry,'' {\em NAW}, vol.~5,
  no.~2, 2005.

\bibitem{jost2008riemannian}
J.~Jost and J.~Jost, {\em Riemannian geometry and geometric analysis},
  vol.~42005.
\newblock Springer, 2008.

\bibitem{HGeiges-ContTopo}
H.~Geiges, {\em An Introduction to Contact Topology}.
\newblock Cambridge University Press, 2008.

\bibitem{Bolsinov2019ChaosAI}
A.~V. Bolsinov, A.~Veselov, and Y.~Ye, ``Chaos and integrability in
  sl(2,r)-geometry,'' {\em arXiv: Geometric Topology}, 2019.

\bibitem{do1992riemannian}
M.~ Do Carmo, {\em ``Riemannian Geometry''}.
\newblock Birkhauser, 1992.



\end{thebibliography}

\end{document}